\documentclass[aps,pre,amsmath,superscriptaddress,amssymb,onecolumn,showpacs, notitlepage]{revtex4-1}     
\usepackage{graphicx}
\usepackage{subfig}
\usepackage{bm}
\usepackage{float}
\usepackage{color}
\usepackage{hyperref}
\usepackage[normalem]{ulem}
\usepackage{enumitem}

\usepackage{cancel}

\usepackage{hyperref}
\definecolor{super-dark-red}{RGB}{165,0,38}
\definecolor{super-dark-blue}{RGB}{8,48,107}
    \definecolor{super-dark-green}{RGB}{0,69,41}
    \definecolor{super-dark-purple}{RGB}{63,0,125}

                \newcommand{\colorred}[1]{{\color{black}#1}}

\begin{document}

\title{Notes on the Hybrid Monte Carlo Method}
	\author{Jeremy C. Palmer}
	\email[Corresponding author:]{jcpalmer@uh.edu}
	\affiliation{Department of Chemical and Biomolecular Engineering, University of Houston, Houston, Texas 77204, USA}

    \author{Amir Haji-Akbari}
    \altaffiliation{Current address: Department of Chemical and Environmental Engineering, Yale University, New Haven, CT 06520, USA}
       \affiliation{Department of Chemical and Biological Engineering, Princeton University, Princeton, New Jersey 08544, USA}
       
    \author{Rakesh S. Singh}
    \affiliation{Department of Chemical and Biological Engineering, Princeton University, Princeton, New Jersey 08544, USA}
    
	\author{Fausto Martelli}
	\author{Roberto Car}
	\affiliation{Department of Chemistry, Princeton University, Princeton, New Jersey 08544, USA}

	\author{Athanassios Z. Panagiotopoulos}
	\author{Pablo G. Debenedetti}
	\email[Corresponding author:]{pdebene@princeton.edu}
	\affiliation{Department of Chemical and Biological Engineering, Princeton University, Princeton, New Jersey 08544, USA}
    \date{\today}

\maketitle

\section{Introduction}\label{intro}
The hybrid Monte Carlo (HMC) method is a widely used molecular simulation technique for computing the equilibrium properties of condensed matter systems \cite{Duane87, Mehlig92,Allen13}. The basic HMC algorithm for sampling from the canonical (NVT) ensemble consists of three steps \cite{Allen13}:
\begin{enumerate}[label=\textbf{Step \arabic*.}, leftmargin=4\parindent]
\item Draw a complete set of initial momenta $\textbf{p} \equiv \left\lbrace\textbf{p}_i\right\rbrace_{i=1}^N$ from the distribution $P(\textbf{p})$. \label{step1}
\item Propagate a microcanonical molecular dynamics (MD) trajectory using a time-reversible and volume-preserving integrator (e.g., the velocity-Verlet algorithm) to take the system from state $\textbf{x}$ to $\textbf{x}'$, where $\textbf{x} \equiv \left\lbrace\textbf{x}_i\right\rbrace_{i=1}^N$ is a complete set of particle coordinates for the system. The length of the trajectory $\Delta t = n_{\text{steps}} \times \delta t$ is specified by the number of integration steps $n_{\text{steps}}$ and time step $\delta t$. \label{step2}
\item Accept or reject the new configuration $\textbf{x}'$ according to the Metropolis-Hastings criterion, \label{step3}
\begin{equation} \label{metropeq}
P_{\text{acc}}(\left\lbrace \textbf{x},\textbf{p} \right\rbrace \rightarrow \left\lbrace \textbf{x}',\textbf{p}' \right\rbrace)  = \text{min}   \left(1, \frac{e^{-\beta U(\textbf{x}')} P(\textbf{p}') }{e^{-\beta U(\textbf{x})}P(\textbf{p})}\right),
\end{equation}
where $U(\textbf{x})$ is the potential energy of the system.
\end{enumerate}

\par  The most common choice for $P(\textbf{p})$ is the Maxwell-Boltzmann (MB) distribution \cite{Duane87, Mehlig92,Allen13},
\begin{equation} \label{pmb}
P_{\text{MB}}^{\text{std}}(\textbf{p}) \propto Ce^{-\beta K(\textbf{p})},
\end{equation}
where $\beta = (k_{B}T)^{-1}$ and $C$ is a temperature specific normalization constant.   The kinetic energy $K(\textbf{p})$ is defined using the standard expression from classical mechanics.  For rigid bodies, for example, $\textbf{p} \equiv  \left\lbrace\textbf{p}_i^\text{com}, \bm{\Omega}_i\right\rbrace_{i=1}^N$, where $\textbf{p}_i^{\text{com}} \equiv m_i \textbf{v}_i^{\text{com}}$, $\textbf{v}_i^{\text{com}}$  is the linear center of mass (`com') velocity for molecule $i$, and $\bm{\omega}_i$ and  $\bm{\Omega}_i$  denote molecule $i$'s angular velocity and angular momentum, respectively.  The $j^\text{th}$ component of molecule $i$'s  angular momentum ($j = 1,2,3$) is $\Omega_j =  {I_{ j,k} \omega_{i,k}}$ $(k=1,2,3)$, where $I_{ j,k}$ denotes  the generic molecule's  ($i$ in this case) inertia tensor. The kinetic energy for rigid bodies is thus given by $K(\textbf{p}) \equiv \frac{1}{2} \sum_{i=1}^{N}  \left[ \textbf{p}_i^{\text{com}} \cdot \textbf{p}_i^{\text{com}}/m_i +  \bm{\omega}_i \cdot \bm{\Omega}_i\right]$.

Insertion of $P_{\text{MB}}(\textbf{p})$ into Eq.\ \ref{metropeq} yields:
\begin{equation}
P_{\text{acc}}^{\text{std}}(\left\lbrace \textbf{x},\textbf{p} \right\rbrace \rightarrow \left\lbrace \textbf{x}',\textbf{p}' \right\rbrace) =  \text{min}\left(1,e^{-\beta \left[ H(\textbf{x}',\textbf{p}') - H(\textbf{x},\textbf{p}) \right]} \right), \label{pacc}
\end{equation}
where $H(\textbf{x}',\textbf{p}') - H(\textbf{x},\textbf{p})\equiv U(\textbf{x}') -U(\textbf{x}) + K(\textbf{p}') - K(\textbf{p})$ is the difference between the classical Hamiltonians $H(\textbf{x},\textbf{p})$ and $H(\textbf{x}',\textbf{p}')$.

\par  The superscript `std' in  Eqs. \ref{pmb} and \ref{pacc} denotes that these are the standard choices for $P(\textbf{p})$ and $P_{\text{acc}}$, which are used in the overwhelming majority of HMC simulation studies of condensed matter systems.  When used together, they ensure that the HMC algorithm will satisfy detailed balance and thus asymptotically sample configurations from the Boltzmann distribution $P_{\text{eq}}(\textbf{x}) \propto e^{-\beta U(\textbf{x})}$.  Sampling can also be performed using other initial momentum distributions that are even functions of $\textbf{p}$ (i.e., $P(\textbf{p}) = P(-\textbf{p})$) \cite{Mehlig92,Allen13}. As we demonstrate below, however, it is generally necessary to modify the acceptance criterion as prescribed by Eq.\ \ref{metropeq} to preserve detailed balance.  Hence if $P(\textbf{p}) \ne P_{\text{MB}}(\textbf{p})$, then using the standard acceptance criterion in Eq. \ref{pacc} will generally result in detailed balance violations and concomitant sampling errors.

\section{Detailed Balance}\label{bg}
\par The detailed balance condition is given by:
\begin{equation}
P_{\text{eq}}(\textbf{x}) \Gamma(\textbf{x} \rightarrow \textbf{x}') = P_{\text{eq}}(\textbf{x}') \Gamma(\textbf{x}' \rightarrow \textbf{x}),\label{bal}
\end{equation} 
where  $P_{\text{eq}}(\textbf{x})$ is the equilibrium distribution and $\Gamma(\textbf{x} \rightarrow \textbf{x}')$ is the Markov transition probability from state $\textbf{x}$ to $\textbf{x}'$.  This condition is sufficient to ensure that $P_{\text{eq}}(\textbf{x})$ will be a stationary state for the Markov processes generated by an MC sampling algorithm. Here, we examine the constraints Eq.\ \ref{bal} imposes on the choice of $P(\textbf{p})$ in the HMC algorithm under the following conditions:
\begin{enumerate}[label=\textbf{(\roman*)}]
\item  The MD trajectories are propagated using a time-reversible and volume-preserving integration scheme. (The latter properties ensures that $d\textbf{x}d\textbf{p} = d\textbf{x}'d\textbf{p}'$; non-volume-preserving integrators may be used in HMC, but a Jacobian factor must be incorporated into the acceptance criterion (Eq.\ \ref{metropeq}) to correct for the resulting compression of phase space \cite{Matubayasi99}). \label{c1}
\item   $P_{\text{eq}}(\textbf{x})$ is given by the Boltzmann distribution ($P_{\text{eq}}(\textbf{x}) \propto e^{-\beta U(\textbf{x})}$) \label{c2}
\item  The trial trajectories are accepted using the standard criterion $P_{\text{acc}}^{\text{std}}$ (Eq.\ \ref{pacc}). \label{c3}
\item  The Hamiltonian $H(\textbf{x},\textbf{p})$ is an even function of $\textbf{p}$ (i.e.,  $H(\textbf{x},-\textbf{p}) = H(\textbf{x},\textbf{p})$) and defined using the standard expression for kinetic energy from classical mechanics. \label{c4}
\item  The function $P(\textbf{p})$ is unspecified, but has the following properties: \label{c5}
\begin{enumerate}[label=\textbf{(\alph*)}]
\item \subitem  $P(\textbf{p}) = P(-\textbf{p})$
\item \subitem $P(\textbf{p})$ is stationary (i.e., the definition of $P(\textbf{p})$ does not change with time). The initial $\textbf{p}$ and final $\textbf{p}'$ momenta are thus both treated as variates from this distribution.  Note that this does not imply  $P(\textbf{p}) = P(\textbf{p}')$, which would suggest ``uniformity''.
\end{enumerate}
\end{enumerate}

For HMC, the forward transition probability from $\textbf{x}$ to $\textbf{x}'$ is given by \cite{Matubayasi99}:
\begin{eqnarray}
\Gamma(\textbf{x} \rightarrow \textbf{x}') = \iint d\textbf{p} d\textbf{p}' P(\textbf{p})g(\left\lbrace \textbf{x},\textbf{p} \right\rbrace \rightarrow \left\lbrace \textbf{x}',\textbf{p}' \right\rbrace) P_{\text{acc}}^{\text{std}}(\left\lbrace \textbf{x},\textbf{p} \right\rbrace \rightarrow \left\lbrace \textbf{x}',\textbf{p}' \right\rbrace) \label{forward0} \\
=\iint d\textbf{p} d\textbf{p}' P(\textbf{p})g(\left\lbrace \textbf{x},\textbf{p} \right\rbrace \rightarrow \left\lbrace \textbf{x}',\textbf{p}' \right\rbrace) \text{min}\left(1,e^{-\beta \left[ H(\textbf{x}',\textbf{p}') - H(\textbf{x},\textbf{p}) \right]} \right), \label{forward1}
\end{eqnarray}
where $g(\left\lbrace \textbf{x},\textbf{p} \right\rbrace \rightarrow \left\lbrace \textbf{x}',\textbf{p}' \right\rbrace) = \delta(\textbf{x}' - \textbf{x}(\Delta t))\delta(\textbf{p}' - \textbf{p}(\Delta t))\delta(\textbf{x} - \textbf{x}(0))\delta(\textbf{p} - \textbf{p}(0))$ and $\delta$ is the Kronecker delta function. The function $g(\left\lbrace \textbf{x},\textbf{p} \right\rbrace \rightarrow \left\lbrace \textbf{x}',\textbf{p}' \right\rbrace)$ gives the probability that an MD trajectory originating from $\lbrace \textbf{x}(0),\textbf{p}(0) \rbrace = \lbrace \textbf{x},\textbf{p} \rbrace$ will end up in state $\lbrace \textbf{x}(\Delta t),\textbf{p}(\Delta t) \rbrace = \lbrace \textbf{x}',\textbf{p}' \rbrace$ after an elapsed time $\Delta t$.

\par  Similarly, the reverse transition probability $\Gamma(\textbf{x}' \rightarrow \textbf{x})$  can be written as \cite{Matubayasi99}:
\begin{eqnarray}
\Gamma(\textbf{x}' \rightarrow \textbf{x}) = \iint d\textbf{p} d\textbf{p}' P(\textbf{p})g(\left\lbrace \textbf{x}',\textbf{p} \right\rbrace \rightarrow \left\lbrace \textbf{x},\textbf{p}' \right\rbrace) P_{\text{acc}}^{\text{std}}(\left\lbrace \textbf{x}',\textbf{p} \right\rbrace \rightarrow \left\lbrace \textbf{x},\textbf{p}' \right\rbrace)  \label{reverse0} \\
=\iint d\textbf{p} d\textbf{p}' P(\textbf{p}) g(\left\lbrace \textbf{x}',\textbf{p} \right\rbrace \rightarrow \left\lbrace \textbf{x},\textbf{p}' \right\rbrace) \text{min}\left(1,e^{-\beta \left[ H(\textbf{x},\textbf{p}') - H(\textbf{x}',\textbf{p}) \right]} \right). \label{reverse1}
\end{eqnarray} 
 In Eqs.\ \ref{reverse0} and  \ref{reverse1} we follow the notation convention used in Ref. \cite{Matubayasi99}, whereby $\textbf{p}$ denotes the generic set of momenta over which integration is performed, and hence we can use $\textbf{p}$ in conjunction with $\textbf{x}'$.  The notation is arbitrary, provided that one defines and uses $\textbf{p}$ and $\textbf{p}'$ consistently and does not confuse these variables in manipulating the equations or making variable substitutions. In particular, because $P$ refers to a generic momentum distribution associated with the initial state of the transition under consideration, one can write, for a $\textbf{x}' \rightarrow \textbf{x}$ transition (Eqs.\ \ref{reverse0} or \ref{reverse1}) $P(\textbf{p})g(\left\lbrace \textbf{x}',\textbf{p} \right\rbrace \rightarrow \left\lbrace \textbf{x},\textbf{p}' \right\rbrace) P_{\text{acc}}^{\text{std}}(\left\lbrace \textbf{x}',\textbf{p} \right\rbrace \rightarrow \left\lbrace \textbf{x},\textbf{p}' \right\rbrace)$ or  $P(\textbf{p}')g(\left\lbrace \textbf{x}',\textbf{p}' \right\rbrace \rightarrow \left\lbrace \textbf{x},\textbf{p} \right\rbrace) P_{\text{acc}}^{\text{std}}(\left\lbrace \textbf{x}',\textbf{p}' \right\rbrace \rightarrow \left\lbrace \textbf{x},\textbf{p} \right\rbrace)$.   Changing (redefining) integration variables \cite{Tuckerman10} $\textbf{p} \rightarrow -\textbf{p}'$ and $\textbf{p}' \rightarrow -\textbf{p}$ and noting that $d\textbf{p} d\textbf{p}' = d\left[ -\textbf{p}' \right]d\left[-\textbf{p}\right]$ yields: 
\begin{eqnarray}
\Gamma(\textbf{x}' \rightarrow \textbf{x}) = \iint d\textbf{p} d\textbf{p}' P(-\textbf{p}') g(\left\lbrace \textbf{x}',-\textbf{p}' \right\rbrace \rightarrow \left\lbrace \textbf{x},-\textbf{p} \right\rbrace) \text{min}\left(1,e^{-\beta \left[ H(\textbf{x},-\textbf{p}) - H(\textbf{x}',-\textbf{p}') \right]} \right). \label{reverse2}
\end{eqnarray} 
For a time-reversible and volume-preserving integrator $ g(\left\lbrace \textbf{x}',-\textbf{p}' \right\rbrace \rightarrow \left\lbrace \textbf{x},-\textbf{p} \right\rbrace ) = g(\left\lbrace \textbf{x},\textbf{p} \right\rbrace \rightarrow \left\lbrace \textbf{x}',\textbf{p}' \right\rbrace)$. 
Additionally, assuming $H(\textbf{x},-\textbf{p}) = H(\textbf{x},\textbf{p})$ and $P(-\textbf{p}) = P(\textbf{p})$, we obtain:
\begin{eqnarray}
\Gamma(\textbf{x}' \rightarrow \textbf{x}) = \iint d\textbf{p} d\textbf{p}' P(\textbf{p}') g(\left\lbrace \textbf{x},\textbf{p} \right\rbrace \rightarrow \left\lbrace \textbf{x}',\textbf{p}' \right\rbrace) \text{min}\left(1,e^{-\beta \left[ H(\textbf{x},\textbf{p}) - H(\textbf{x}',\textbf{p}') \right]} \right). \label{reverse3}
\end{eqnarray} 
Invoking the identity $\text{min}\left(1,e^{-\beta \left[ H(\textbf{x},\textbf{p}) - H(\textbf{x}',\textbf{p}') \right]} \right) = \frac{e^{-\beta H(\textbf{p},\textbf{x})}}{e^{-\beta H(\textbf{p}',\textbf{x}')}}\text{min}\left(1,e^{-\beta \left[ H(\textbf{x}',\textbf{p}') - H(\textbf{x},\textbf{p}) \right]} \right)$:
\begin{eqnarray}
\Gamma(\textbf{x}' \rightarrow \textbf{x}) = \iint d\textbf{p} d\textbf{p}' P(\textbf{p}') g(\left\lbrace \textbf{x},\textbf{p} \right\rbrace \rightarrow \left\lbrace \textbf{x}',\textbf{p}' \right\rbrace) \frac{e^{-\beta H(\textbf{p},\textbf{x})}}{e^{-\beta H(\textbf{p}',\textbf{x}')}}\text{min}\left(1,e^{-\beta \left[ H(\textbf{x}',\textbf{p}') - H(\textbf{x},\textbf{p}) \right]} \right) \label{reverse4}\\
=\iint d\textbf{p} d\textbf{p}' P(\textbf{p}) g(\left\lbrace \textbf{x},\textbf{p} \right\rbrace \rightarrow \left\lbrace \textbf{x}',\textbf{p}' \right\rbrace) \frac{P(\textbf{p}')}{P(\textbf{p})} \frac{e^{-\beta U(\textbf{x})}e^{-\beta K(\textbf{p})}}{e^{-\beta U(\textbf{x}')}e^{-\beta K(\textbf{p}')}}\text{min}\left(1,e^{-\beta \left[ H(\textbf{x}',\textbf{p}') - H(\textbf{x},\textbf{p}) \right]} \right) \label{reverse5}\\
=  \frac{e^{-\beta U(\textbf{x})}}{e^{-\beta U(\textbf{x}')}} \iint d\textbf{p} d\textbf{p}' P(\textbf{p}) g(\left\lbrace \textbf{x},\textbf{p} \right\rbrace \rightarrow \left\lbrace \textbf{x}',\textbf{p}' \right\rbrace) \boxed{\frac{P(\textbf{p}')e^{-\beta K(\textbf{p})}}{P(\textbf{p})e^{-\beta K(\textbf{p}')}}}\text{min}\left(1,e^{-\beta \left[ H(\textbf{x}',\textbf{p}') - H(\textbf{x},\textbf{p}) \right]} \right).\label{reverse6}
\end{eqnarray}
Comparison with Eq. \ref{forward1} reveals that the integrands differ by the boxed term in Eq. \ref{reverse6}.  To satisfy detailed balance (Eq.\ \ref{bal}), the choice of $P(\textbf{p})$ must therefore either ensure that the boxed term in  Eq. \ref{reverse6} is always unity or has no effect on the value of the integral when the integration over $\textbf{p}$ and $\textbf{p}'$ is carried out.   It is clear from our derivation that neither of these criteria will be automatically satisfied by an arbitrary distribution $P(\textbf{p})$, even if it has the properties listed in point \ref{c5} above. The former criterion is satisfied, however, by the Maxwell-Boltzmann distribution $P(\textbf{p}) = P^{\text{std}}_{\text{MB}}(\textbf{p})$ (Eq.\ \ref{pmb}). Inserting the MB distribution into Eqs.\ \ref{forward1} and \ref{reverse6}, we find that the boxed term in Eq.\ \ref{reverse6} is identically equal to 1. Subsequent inspection of Eqs.\ \ref{forward1} and \ref{reverse6} reveals that $\Gamma(\textbf{x} \rightarrow \textbf{x}') = \frac{e^{-\beta U(\textbf{x}')}}{e^{-\beta U(\textbf{x})}} \Gamma(\textbf{x}' \rightarrow \textbf{x}) $, which is the detailed balance condition (Eq.\ \ref{bal}).  By contrast, identifying a distribution that unambiguously satisfies only the latter criterion is extremely challenging, as the integrals in Eq. \ref{forward1} and \ref{reverse6} cannot usually be evaluated analytically.  Other choices for $P(\textbf{p})$ are certainly possible, but HMC acceptance criterion in Eqs.\ \ref{forward1} and \ref{reverse6} must be redefined as prescribed by Eq.\ \ref{metropeq} to avoid violating detailed balance. 

\section{Detailed Balance for the One-dimensional Harmonic Oscillator}\label{MBproof}

\par The analysis above demonstrates that choosing $P(\textbf{p})=P^{\text{std}}_{\text{MB}}(\textbf{p})$ is generally required to satisfy detailed balance when HMC sampling is performed using the standard acceptance criterion $P_{\text{acc}}^{\text{std}}$ (Eq.\ \ref{pacc}). \colorred{Here, we show analytically for a simple model system and velocity Verlet integrator that detailed balance can \emph{only} be satisfied by $P(\textbf{p})=P^{\text{std}}_{\text{MB}}(\textbf{p})$ when $P_{\text{acc}}=P_{\text{acc}}^{\text{std}}$. For this case, the condition that $P(\textbf{p})=P^{\text{std}}_{\text{MB}}(\textbf{p})$ is thus not only sufficient but also necessary.}  Consider the simple case of applying the HMC algorithm to sample the configurations of a particle of mass $m$ in a one-dimensional harmonic well described by the potential energy function $U(x) = \frac{1}{2} k x^2$. Further, suppose that we define the kinetic energy using the standard expression $K(p) \equiv \frac{p^2}{2m}$ and consider HMC moves that consist of only a \emph{single} molecular dynamics integration step propagated using the time reversible and volume-preserving velocity Verlet scheme.  Under these conditions, $\left\lbrace x', p'\right\rbrace$ is related to $\left\lbrace x, p\right\rbrace$ via:
\begin{eqnarray}
x' = x + \frac{p}{m}\delta t + \frac{f(x)}{2m}\delta t^2 \\
p' = p + \frac{1}{2} \left[ f(x') + f(x) \right] \delta t,
\end{eqnarray} 
where $\delta t$ is the integration time step and $f(x) = -\frac{dU(x)}{dx} = -kx $ is the force acting on the particle. The detailed balance condition (Eq.\ \ref{bal}) must be satisfied for all $x$ and $x'$. Thus we are free to choose $x$ and $x'$ arbitrarily to simplify our analysis.   Letting $x=0$ and $x' \ne 0$, and noting that $f(0)=0$ and $f(x') = -kx'$, we obtain:
\begin{eqnarray}
x' = \frac{p}{m}\delta t \\
p' = p - \frac{1}{2} kx' \delta t  = p(1-\frac{1}{2m}k\delta t^2).
\end{eqnarray}
These equations show that for a given choice of the sampling parameter $\delta t$, only $p = \frac{x'm}{\delta t}$ will lead to an attempted HMC move from  $\left\lbrace x,p \right\rbrace \rightarrow \left\lbrace x',p' \right\rbrace$.  Similarly, they also reveal that $p' = \frac{x'm}{\delta t} - \frac{1}{2} kx' \delta t$ is unique for a given $\delta t$. 
Therefore $g$ will be given by:
\begin{eqnarray} \label{gho}
g(\left\lbrace x, p\right\rbrace \rightarrow \left\lbrace x', p'\right\rbrace)=\delta\left(p-\frac{x'm}{\delta t}\right)
\delta\left(p'-\frac{x'm}{\delta t}+\frac12kx'\delta t\right).
\end{eqnarray}
Additionally, we find that:
\begin{eqnarray} 
H(x,p) =  \frac{x'^2m}{2\delta t^2}  \label{hho0} \\
H(x',p') =   \frac{x'^2m}{2\delta t^2} + \frac{1}{8m}k^2x'^2 \delta t^2 \label{hho1}
\end{eqnarray}

Using these expressions, we can now analyze the conditions under which the detailed balance (Eq.\ \ref{bal}) will be satisfied. For the selected $x$ and $x'$, the detailed balance condition (Eq.\ \ref{bal}) reduces to:
\begin{eqnarray} 
 \Gamma(x \rightarrow x') = \frac{e^{-\beta U(x')}}{e^{-\beta U(x)}} \Gamma(x' \rightarrow x) \\
 = {e^{-\frac{\beta}{2}kx'^2 }} \Gamma(x' \rightarrow x) \label{balho},
\end{eqnarray} 
where we have substituted $U(x) = 0 $ and $U(x') = \frac{1}{2}kx'^2$.  Inserting Eqs.\ \ref{gho}, \ref{hho0}, and \ref{hho1} into the expression for the forward transition probability given by Eq.\ \ref{forward1}, we obtain:
\begin{eqnarray} 
\Gamma(x \rightarrow x') =\iint dp dp' P(p)g(\left\lbrace x, p \right\rbrace \rightarrow \left\lbrace x',p' \right\rbrace) \text{min}\left(1,e^{-\beta \left[ H(x',p') - H(x,p) \right]} \right) \\
=\iint dp dp' P(p)\delta\left(p-\frac{x'm}{\delta t}\right) \delta\left(p'-\frac{x'm}{\delta t}+\frac12kx'\delta t\right) \text{min}\left(1,e^{-\beta \left[ \frac{x'^2m}{2\delta t^2} + \frac{1}{8m}k^2x'^2 \delta t^2  - \frac{x'^2m}{2\delta t^2} \right]} \right) \\
= P\left(\frac{x'm}{\delta t}\right) \text{min}\left(1,e^{ -\frac{\beta}{8m}k^2x'^2 \delta t^2 } \right) \\
= P\left(\frac{x'm}{\delta t}\right) e^{ -\frac{\beta}{8m}k^2x'^2 \delta t^2 }, \label{forwardho}
\end{eqnarray} 
where the last equality holds because $\frac{\beta}{8m}k^2x'^2 \delta t^2  \ge 0$.  The reverse transition probability can also be evaluated in a similar fashion from Eq.\ \ref{reverse4}:
\begin{eqnarray} 
\Gamma(x' \rightarrow x) = \iint dp dp' P(p') g(\left\lbrace x,p \right\rbrace \rightarrow \left\lbrace x',p' \right\rbrace) \frac{e^{-\beta H(p,x)}}{e^{-\beta H(p',x')}}\text{min}\left(1,e^{-\beta \left[ H(x',p') - H(x,p) \right]} \right) \\
= \iint dp dp' P(p') \delta\left(p-\frac{x'm}{\delta t}\right) \delta\left(p'-\frac{x'm}{\delta t}+\frac12kx'\delta t\right)  \frac{e^{-\beta H(p,x)}}{e^{-\beta H(p',x')}}\text{min}\left(1,e^{-\beta \left[ H(x',p') - H(x,p) \right]} \right) \\
= P\left(\frac{x'm}{\delta t}-\frac12kx'\delta t \right)  e^{\frac{\beta}{8m}k^2x'^2 \delta t^2 }\text{min}\left(1,e^{-\frac{\beta}{8m}k^2x'^2 \delta t^2} \right)\\
= P\left(\frac{x'm}{\delta t}-\frac12kx'\delta t \right)  e^{\frac{\beta}{8m}k^2x'^2 \delta t^2 }e^{-\frac{\beta}{8m}k^2x'^2 \delta t^2} \\
= P\left(\frac{x'm}{\delta t}-\frac12kx'\delta t \right) \label{reverseho}
\end{eqnarray}

Substituting Eqs.\ \ref{forwardho} and \ref{reverseho} into Eq.\ \ref{balho}, we find that for detailed balance to hold, we need to have:
\begin{eqnarray}
P\left(\frac{x'm}{\delta t}\right)e^{-\frac{\beta}{8m}k^2x'^2\delta t^2}&=& 
 e^{-\frac{\beta}2kx'^2}P\left(\frac{x'm}{\delta t}-\frac12kx'\delta t\right)
\end{eqnarray}
which is clearly not satisfied for an arbitrary $P(\cdot)$. Letting $\phi(p):=\ln P(p)+\beta p^2/2m$, we obtain:
\begin{eqnarray}
 \ln P\left(\frac{x'm}{\delta t}\right) - \frac{\beta}{8m}k^2x'^2\delta t^2 &=&\ln P\left(\frac{x'm}{\delta t}-\frac12kx'\delta t\right) - \frac{\beta}2kx'^2\notag\\
 \phi\left(\frac{x'm}{\delta t}\right)-\frac{\beta m x'^2}{2\delta t^2} - \frac{\beta}{8m}k^2x'^2\delta t^2&=&  \phi\left(\frac{x'm}{\delta t}-\frac12kx'\delta t\right) -\frac{\beta}{2m}\left(\frac{x'm}{\delta t}-\frac12kx'\delta t\right)^2- \frac{\beta}2kx'^2 \notag\\
\phi\left(\frac{x'm}{\delta t}\right)-\cancel{\frac{\beta m x'^2}{2\delta t^2}} - \cancel{\frac{\beta}{8m}k^2x'^2\delta t^2} &=&    \phi\left(\frac{x'm}{\delta t}-\frac12kx'\delta t\right) -\cancel{\frac{\beta m x'^2}{2\delta t^2}} + \cancel{\frac{\beta}2kx'^2}-\cancel{\frac{\beta}{8m}k^2x'^2\delta t^2} - \cancel{\frac{\beta}2kx'^2}\notag\\
 \phi\left(\frac{x'm}{\delta t}\right) &=&  \phi\left(\frac{x'm}{\delta t}-\frac12kx'\delta t\right)\label{eqphi}
\end{eqnarray} 
which can be mathematically stated as $\phi(bx')=\phi(ax')$ with $b=m/\delta t, a=m/\delta t-k\delta t/2$.   Because $\phi(bx')=\phi(ax')$ for all $x'$, 
\begin{eqnarray}
\phi(b^2x') = \phi\left[ b(bx') \right] = \phi\left[ a(bx') \right] = \phi\left[ b(ax') \right] =\phi\left[ a(ax') \right]  = \phi( a^2x'),
\end{eqnarray}
or more generally,
\begin{eqnarray}
\phi(b^nx')  = \phi(a^nx')
\end{eqnarray}
for $n\in\mathbb{Z}$. 
Similarly,
\begin{eqnarray}\label{phin}
\phi(x')  = \phi\left[b^n \frac{x'}{b^n} \right] = \phi\left[a^n \frac{x'}{b^n} \right] =\phi\left[\left(\frac{a}{b}\right)^nx'\right].
\end{eqnarray}
Without loss of generality, suppose $a<b$. Because Eq.\ \ref{phin} is valid for any $n\in\mathbb{Z}$,  it must also hold in the limit that $n\rightarrow\infty$. Taking this limit, we find that:
$$\phi(x') = \phi\left[\left(\frac{a}{b}\right)^nx'\right] = \lim_{n\rightarrow\infty} \phi\left[\left(\frac{a}{b}\right)^nx'\right] = \phi(0)=\text{const.}\notag$$
Thus,  in order for Eq.\ \ref{eqphi} to hold for \emph{any} $x'$, $\phi(p)$ needs to be constant:
\begin{eqnarray}
\phi(p)=\ln P(p)+\beta p^2/2m  = \text{const.},
\end{eqnarray}
or equivalently,
\begin{eqnarray}
P(p) \propto e^{-\beta \frac{p^2}{2m}} = e^{-\beta K(p)},
\end{eqnarray}
which is the Maxwell-Boltzmann distribution.  Hence, this derivation proves for the 1-D harmonic oscillator model system that for detailed balance to hold when the standard HMC acceptance criterion  (Eq.\ \ref{pacc}) is used in conjunction with $K(p) \equiv \frac{p^2}{2m}$, $P(p)$ can \emph{only} have the Maxwell-Boltzmann distribution.  

\section{A Stationary and even momentum distribution is not sufficient for detailed balance}\label{LJ}
\par
\begin{figure}[h]
\hspace{20px}
\includegraphics [width =0.25 \linewidth ]{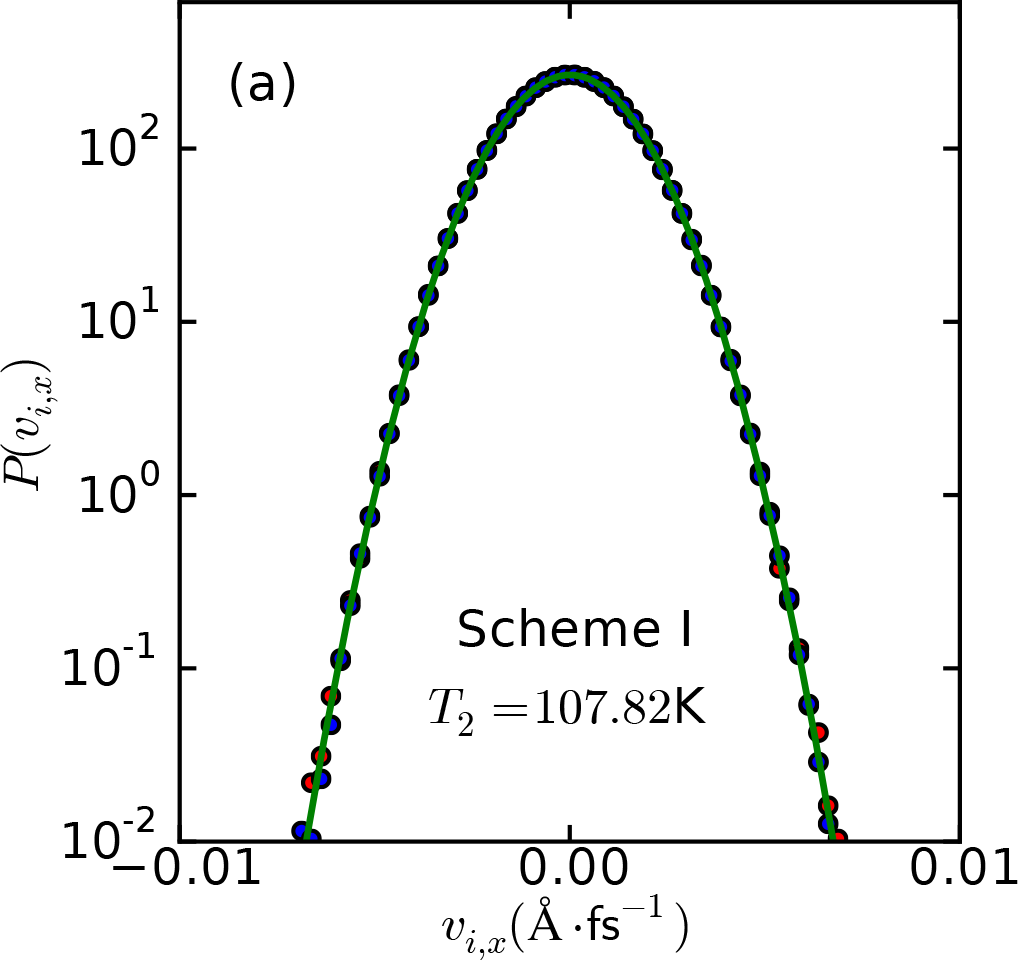}
\includegraphics [width =0.25\linewidth ]{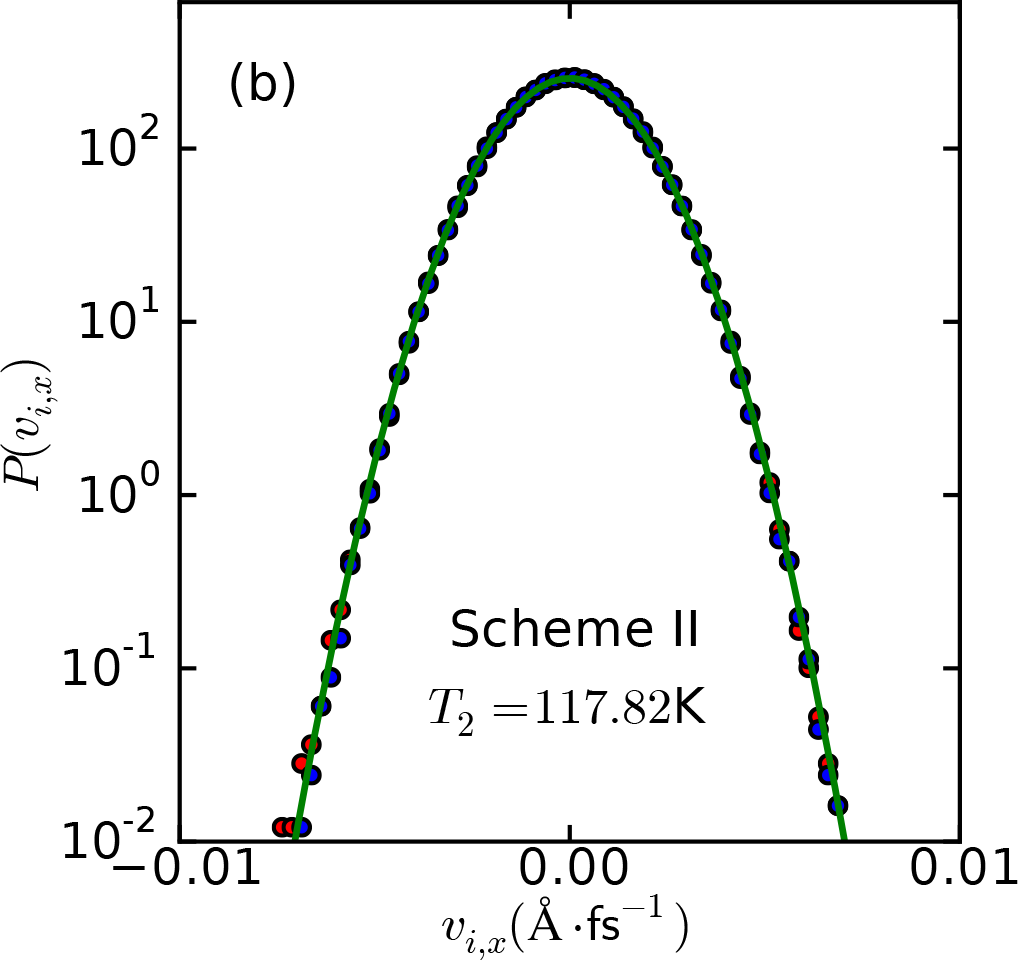}
\hspace{20px}
\includegraphics [width =0.25 \linewidth  ]{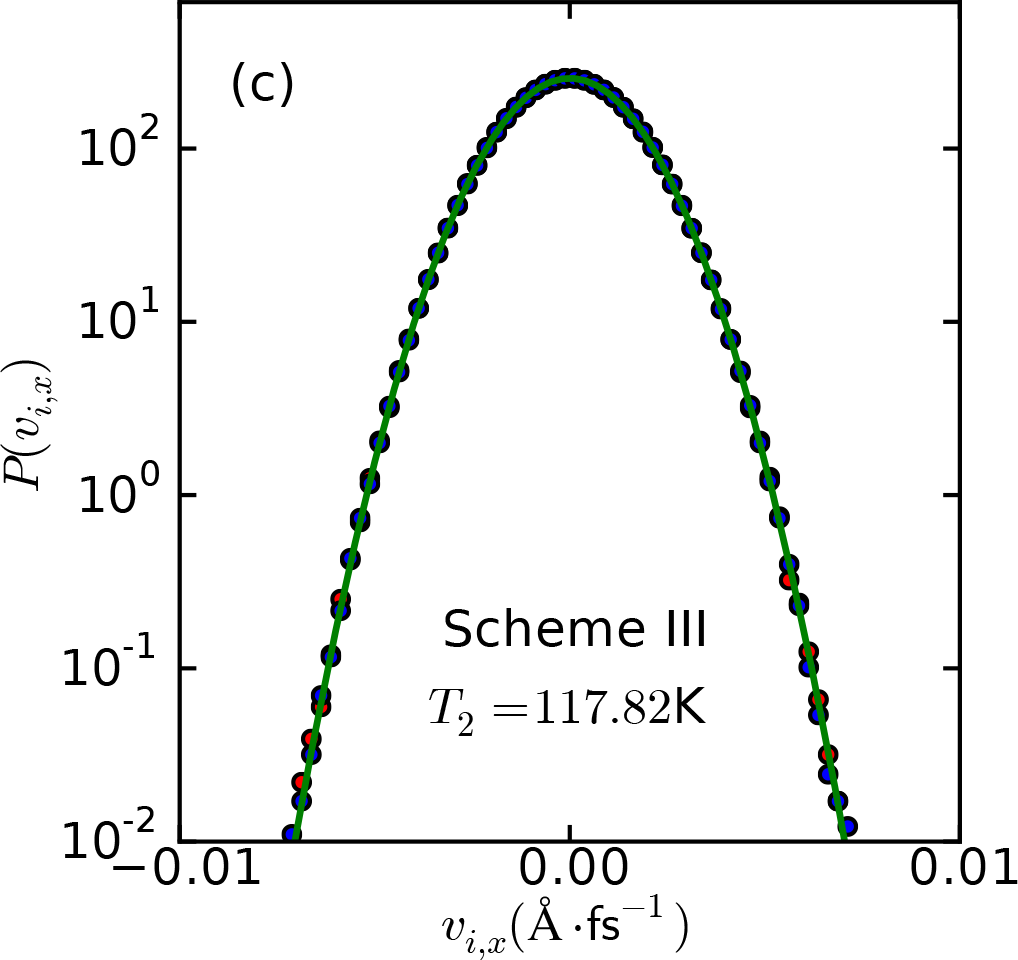}
\includegraphics [ width =0.25 \linewidth ]{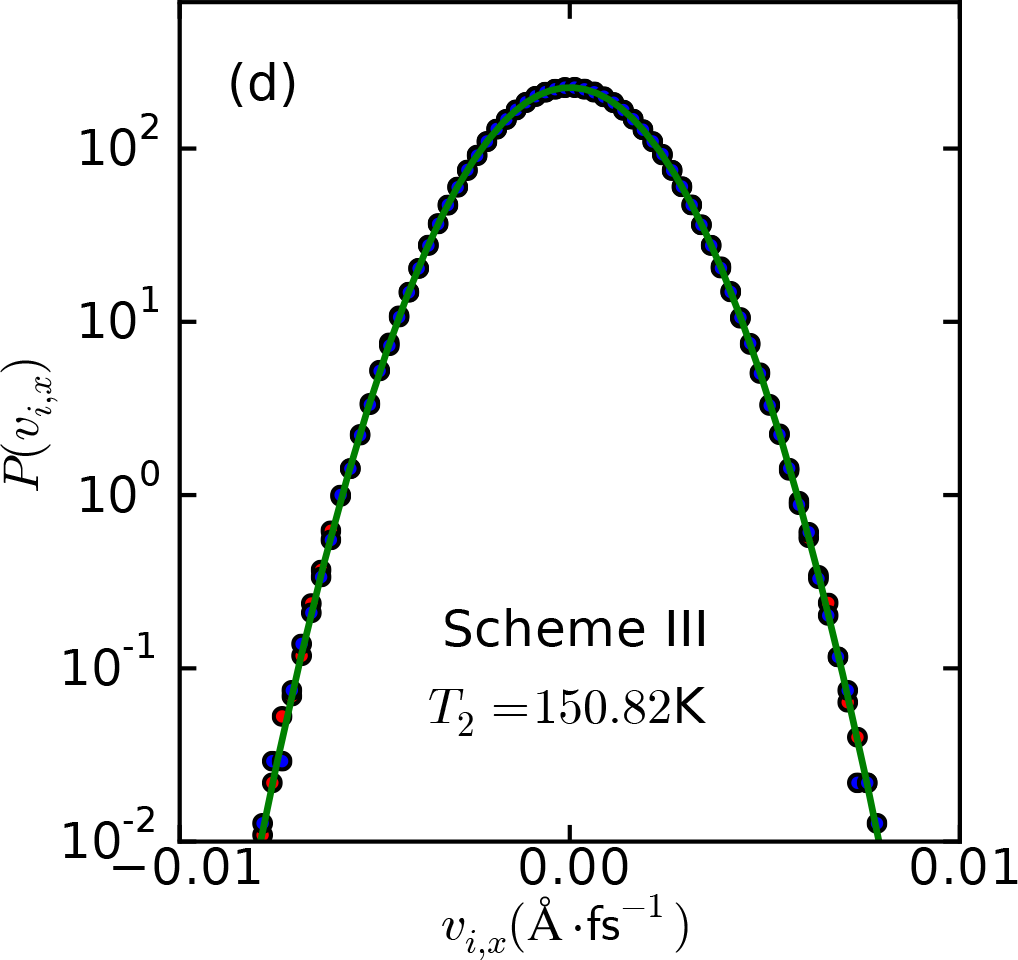}
\includegraphics [width =0.25 \linewidth  ]{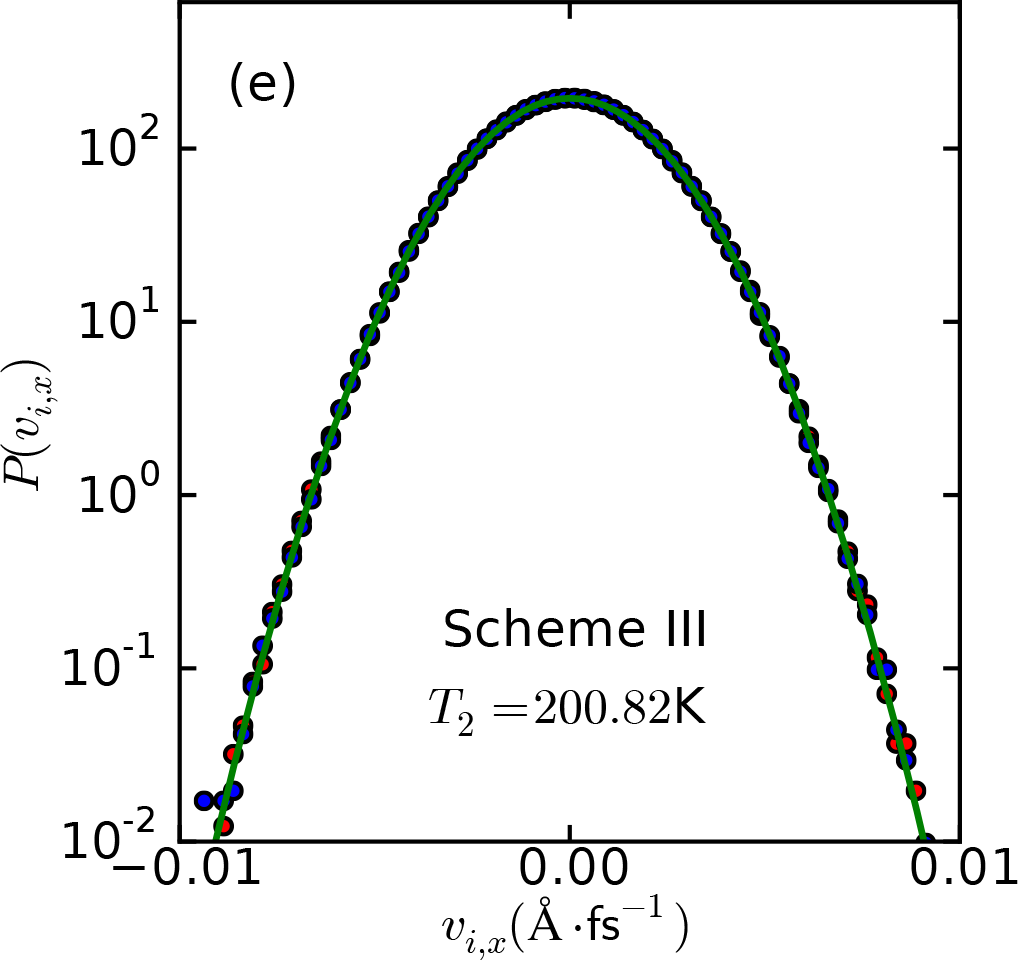}
\caption {\label{LJvel} The proposed  (red) and accepted  (blue) velocity distributions are statistically indistinguishable for all sampling schemes and choices of $T_2$.  Thus, the velocity distribution is stationary regardless of whether detailed balance is satisfied by the algorithm.  The green lines show the Maxwell-Boltzmann distribution at the indicated temperature $T_2$.   Scheme III does not satisfy detailed balance and hence does not yield the correct equation of state for the Lennard-Jones fluid (Table \ref{LJeos}), even though the velocity distribution is an even function and stationary. These two criteria are therefore \emph{not sufficient} to ensure detailed balance}
\end{figure}

 The derivations outlined in Secs.\ \ref{bg}  and \ref{MBproof} assume that $P(\textbf{p})$ is even (i.e., $P(\textbf{p})=P(-\textbf{p})$) and stationary (i.e., the initial momenta $\textbf{p}$ and final momenta $\textbf{p}'$ are treated as variates from the same distribution, $P(\textbf{p})$).  As demonstrated by our analysis, however, these properties are not sufficient to satisfy detailed balance. This fact can also be shown numerically by performing HMC simulations of the Lennard-Jones model for argon ($\sigma_{\text{Ar}} = 0.3405$ nm, $\epsilon_{\text{Ar}}/k_B = 119.8 $ K), in which the potential is truncated at $ 3 \times \sigma_{\text{Ar}}$ and standard long-range tail corrections are applied.  We consider three HMC schemes in which the target sampling temperature is specified by $\beta_1 = (k_BT_1)^{-1}$:
\begin{itemize}
\item \textbf{Scheme I:}  HMC using $P(\textbf{p}) \propto e^{-\beta_2K(\textbf{p})}$ and $P^{\text{std}}_{\text{acc}} =  \text{min}\left(1, \frac{e^{-\beta_1 U(\textbf{x}')}e^{-\beta_1 K(\textbf{p}')}}{e^{-\beta_1 U(\textbf{x})}e^{-\beta_1 K(\textbf{p})}} \right)$, where $\beta_2 = \beta_1$. This scheme corresponds to the standard HMC algorithm, where the initial momenta are drawn from the Maxwell-Boltzmann distribution at the target sampling temperature $\beta_1 = \beta_2 = (k_BT_1)^{-1}$.  As proved in Sec.\ \ref{bg}, this scheme satisfies detailed balance.
  
\item \textbf{Scheme II:} HMC using $P(\textbf{p}) \propto e^{-\beta_2K(\textbf{p})}$ and $P^{\text{mod}}_{\text{acc}} =  \text{min}\left(1, \frac{e^{-\beta_1 U(\textbf{x}')}e^{-\beta_2 K(\textbf{p}')}}{e^{-\beta_1 U(\textbf{x})}e^{-\beta_2 K(\textbf{p})}} \right)$, where $\beta_2 < \beta_1$. In this scheme,  the initial momenta are drawn from a Maxwell-Boltzmann distribution corresponding to an artificially high temperature $\beta_2 = (k_BT_2)^{-1}$, but the acceptance criterion has been modified such that $P^{\text{mod}} _{\text{acc}}= \text{min}   \left(1, \frac{e^{-\beta_1 U(\textbf{x}')} P(\textbf{p}') }{e^{-\beta_1U(\textbf{x})}P(\textbf{p})}\right)$, as prescribed by Eq.\ \ref{metropeq}. This modified acceptance criterion is sufficient to ensure that detailed balance is satisfied for all choices of $\beta_1$ and  $\beta_2$.
  
\item \textbf{Scheme III:} HMC using $P(\textbf{p}) \propto e^{-\beta_2K(\textbf{p})}$ and $P^{\text{std}}_{\text{acc}} =  \text{min}\left(1, \frac{e^{-\beta_1 U(\textbf{x}')}e^{-\beta_1 K(\textbf{p}')}}{e^{-\beta_1 U(\textbf{x})}e^{-\beta_1 K(\textbf{p})}} \right)$, where $\beta_2 < \beta_1$. In this scheme,  the initial momenta are drawn from a Maxwell-Boltzmann distribution corresponding to an artificially high temperature $\beta_2 = (k_BT_2)^{-1}$, but no modifications are made to the standard acceptance criterion.  \textbf{As proved in Sec.\ \ref{MBproof}, this sampling scheme, does not satisfy detailed balance, even though $P(\textbf{p})$ is an even function and stationary (Fig.\ \ref{LJvel})} 
\end{itemize}

\par We simulate $N=500$ particles at a fixed density of 1.3778 g/cm$^3$ and choose $T_1 =107.82 K$ as our target sampling temperature. In terms of Lennard-Jones units, this corresponds to $\rho^* = 0.820$ and $T^* = 0.9$. In each case, sampling is performed for $\sim 5 \times 10^5$ HMC moves, where each move consists of 10 molecular dynamics integration steps using a 30 fs time step. 
\par As expected, Scheme I correctly predicts the average potential energy, as illustrated by the excellent agreement with benchmark data from the National Institute for Science and Technology for the Lennard-Jones fluid (Table \ref{LJeos}).  As expected, it also satisfies the normalization condition $\langle e^{-\beta \Delta H} \rangle = 1$ (Table \ref{LJeos}).  This rigorous statistical mechanical relationship must hold for all valid HMC sampling schemes. Thus it provides a convenient consistency check for detecting sampling errors associated with detailed balance violations.  In the absence of sampling errors, we expect that $\left| \langle e^{-\beta \Delta H} \rangle -1 \right| \times \sigma^{-1} \lesssim 1 $, where $\sigma$ is the estimated uncertainty.  Accordingly, we find that Scheme I satisfies this expectation. Similarly, Scheme II correctly predicts the average potential energy and obeys the normalization condition.  Even though the initial velocities in Scheme II are drawn from the Maxwell-Boltzmann distribution at an artificially high temperature $T_2 = 117.82$ K, the modified HMC acceptance ensures that detailed balance is satisfied and that the correct result is recovered. By contrast, Scheme III does not satisfy detailed balance and hence fails to obey the normalization condition and correctly predict the equation of state (Table \ref{LJeos}).  The sampling errors produced by Scheme III become more pronounced as the velocity generation temperature $T_2$ increases from 117.82 to 200.82 K.

\begin{table}[h]
\caption{\label{LJeos} Equation of state data for Lennard-Jones argon at a target sampling temperature $T_1 = 107.82$ K}
\begin{tabular}{lcc}
\hline
  & $\langle U \rangle/N$ (kJ mol$^{-1}$) &  $ \left| \langle e^{-\beta \Delta H} \rangle -1 \right| \times \sigma^{-1} $  \\
\hline
NIST Ref. Data\footnote{Benchmark data from the \href{https://mmlapps.nist.gov/srs/LJ_PURE/mc.htm}{National Institute of Standards and Technology \url{https://mmlapps.nist.gov/srs/LJ_PURE/mc.htm}}, which has been converted from Lennard-Jones units}  & -5.7230(7)\footnote{Number in parentheses denotes uncertainty in the last significant digit}  & --   \\
Scheme I ($T_1 = 107.82$ K) &  -5.7231(1)  & 0.68  \\
Scheme II ($T_1 = 107.82$ K; $T_2 = 117.82$ K)  & -5.7231(2) & 0.47  \\
Scheme III ($T_1 = 107.82$ K; $T_2 = 117.82$ K)  & -5.6444(1) & 3.5  \\
Scheme III ($T_1 = 107.82$ K; $T_2 = 150.82$ K)  & -5.3998(2) & 21.4  \\
Scheme III ($T_1 = 107.82$ K; $T_2 = 200.82$ K)  & -5.0623(2) & 12.5  \\
\hline
\end{tabular}
\end{table}

\par  In each sampling scheme, $P(\textbf{p})$ is an even function of $\textbf{p}$ by construction. Additionally, $P(\textbf{p})$ is stationary under each sampling scheme, regardless of whether detailed balance is satisfied by the algorithm (Fig.\ \ref{LJvel}). Thus, in agreement with our proof in Sec.\ \ref{MBproof}, these results demonstrate that stationarity and evenness of $P(\textbf{p})$  are \emph{not sufficient} conditions for detailed balance. If one wishes to use the standard HMC acceptance criterion  and definition of kinetic energy given in Sec.\ \ref{intro},  \colorred{initial velocities should be drawn from the Maxwell-Boltzmann distribution at the target sampling temperature to satisfy detailed balance  (see  Sec.\ \ref{MBproof})}. As Scheme II illustrates, detailed balance can be satisfied using other choices for $P(\textbf{p})$, if the HMC acceptance criterion is modified appropriately.   

\section{HMC using partial momentum updates}\label{corrp}

\par The HMC algorithm outlined in Sec.\ \ref{intro} assumes that a completely new set of initial momenta are drawn from $P(\textbf{p})$ in \ref{step1}  It also possible to perform HMC using correlated samples from $P(\textbf{p})$ that are generated by partially refreshing the momenta from the end of the previous MC step. Wagoner and Pande \cite{Wagoner12} have rigorously proved, however, that the sign of the momenta must be changed (negated) either upon acceptance or rejection of each HMC move for this scheme to satisfy balance (i.e., leave $P_{\text{eq}}(\textbf{x}) \propto e^{-\beta U(\textbf{x})}$ stationary) and thus sample from the correct equilibrium distribution. Indeed, numerical studies have shown that omitting this important step results in detectable sampling errors \cite{Akhmatskaya09}.


\vspace{0.5in}
\begin{thebibliography}{0}
\bibitem{Duane87} S. Duane, A. D. Kennedy, B. J. Pendleton, and D. Roweth, Phys. Lett. B {\bf 195}, 216-222 (1987).
\bibitem{Mehlig92} B. Mehlig, D. W. Heermann, and B. M. Forest, Phys. Rev. B {\bf 45}, 679-685 (1992).
\bibitem{Allen13} M. Allen and D. Quigley,  Mol. Phys. {\bf 111}, 3442-3447 (2013).
\bibitem{Matubayasi99} N. Matubayasi and M. Nakahara,  J. Chem. Phys. {\bf 110}, 3291(1999).
\bibitem{Tuckerman10}  M. Tuckerman, Statistical Mechanics: Theory and Molecular Simulation (Oxford University Press, 2010).
\bibitem{Wagoner12}  J. A. Wagoner and V. S. Pande,  J. Chem. Phys. {\bf 137}, 214105 (2012).
\bibitem{Akhmatskaya09} E. Akhmatskaya, N. Bou-Rabee, and S. Reich, J. Comp. Phys. {\bf 19}, 7492-7496 (2009).
\end{thebibliography}
\end{document}